\documentclass{aa}  
\usepackage{graphicx}
\usepackage{subfigure}
\usepackage{txfonts}
\usepackage{xcolor}

\usepackage[allcolors=blue]{hyperref}
\begin{document}

   \title{ Searching for the nature of stars with debris disks and planets}

   \authorrunning{de la Reza et al. }
   \author{R. de la Reza, 
          \inst{1}\fnmsep\thanks{ramirodelareza@yahoo.com}
          C. Chavero,
          \inst{2,3}
           S. Roca-F\`abrega,
          \inst{4,5}
           F. Llorente de Andr\'es,
          \inst{6,7}
          P.  Cruz,
            \inst{7}
          \and
          C. Cifuentes
          \inst{7}
        }

 \institute{ Observatório Nacional,  Rua General Jos\'e Cristino 77, 28921-400 São Cristovão, Rio de Janeiro, RJ, Brasil
     \and
        Observatorio Astron\'omico de C\'ordoba, Universidad Nacional de C\'ordoba, Laprida 854, 5000 C\'ordoba 
      \and
        Consejo Nacional de Investigaciones Científicas y Técnicas (CONICET), Godoy Cruz 2290, Ciudad Autónoma de Buenos Aires, Argentina
      \and
       Departamento de F\'isica de la Tierra y Astrof\'isica, UCM, and IPARCOS, Facultad de Ciencias F\'isicas, Plaza Ciencias, 1, Madrid, E-28040, Spain
      \and
        Instituto de Astronomía, Universidad Nacional Autónoma de México, Apartado Postal 106, C. P. 22800, Ensenada, B. C., Mexico
      \and
        Ateneo de Almagro, Secci\'on de Ciencia y Tecnolog\'ia, 13270 Almagro, Spain
       \and
       Centro de Astrobiología (CAB), CSIC-INTA, Camino Bajo del Castillo s/n, campus ESAC, 28692, Villanueva de la Cañada, Madrid, Spain
       \
       }


 \abstract{
 The  nature of the few known solar-mass stars simultaneously containing debris
disks and planets  remains an open question. A number of works have shown that this property appears to be independent of planetary masses as well  as of stellar age, but possible correlations with stellar kinematics and metallicity have not been investigated. In this paper, we show that the majority of known stars containing both debris disks and planets belong to the metal-enriched Galactic thin disk. The few exceptions are stars that seem to be born in the star formation peak occurring in times of thick disk formation (i.e., HD\,10700, HD\,20794, and HD\,40307), that is, between 11 and 8 Gyr. The mass of the dusty disk of these three old stars measured at 70\,$\mu$m is very small -- in fact, it is lower than that of the Kuiper belt of our Solar system by several orders of magnitude. These results are not surprising, as they remain within the values expected for the stellar disk evolution of such primitive stars. In parallel, we found another six thick-disk stars containing only debris disks or planets. These results enable us to establish a correlation between stellar metallicity and the mass of the dust disk modulated by the different formation epochs of the thick and thin Galactic disks. 
}
  

 \maketitle
%

\section{Introduction}\label{sec:intro}
 
In the initial evolutionary stages of solar-type stars, they are usually surrounded by gas and dust disks characterized by a range of properties that are referred to as protoplanetary disks \citep{Williams2011}.
These are structures that contain mostly small dust grains and gas and they embody the first stages of planetary formation. Over time, this first generation of disks condense into planetesimals, which aggregate to produce minor mass planets and comets to finally produce massive solid structures that become the core of giant
planets after capturing the ambient gas \citep{Pollack1996}. This first stage
lasts from 3 to 10\,Myr, after the first fusion reactions start in the central star.  This is this the time needed for the stellar winds produced by the newborn star to remove the gas component from its surroundings \citep{Williams2011}. 

After this period, a new kind of disk can emerge from the dust produced in the planetesimal collisions. These disks are formed of asteroids and small planetesimals with sizes ranging from a few millimetres to several tens of kilometres \citep{Wyatt2008}. The lifetime of a debris disk may be as long as that of its host star \citep{Williams2011}.

Stars containing debris disks and confirmed planets (DDP) are peculiar objects. In the literature, there are  only about 30 stars that are known with confirmed observations of planets (SVO - {\tt VOMultiCatalog\_Interface}\footnote{http://svo2.cab.inta-csic.es}). It is not clear why the number of these objects is so small given the hundreds of observed stars with debris disks without confirmed planets (DD) or the thousands of stars containing only confirmed planets (CP). Some authors have looked for correlations between the existence of DDPs and the mass of the exoplanets therein, as well as the age of the stellar system, but without success \citep[e.g.,][]{Marshall2014,Montesinos2016}. In fact, DDP stars have been shown to contain a broad range of planetary masses, from terrestrial mass type planets up to giants with several Jupiter masses. In addition, they have been found to exhibit stellar ages that range from the times of maximum star formation in the formation of the Galactic thick disk 8--11 Gyr ago up to the recent formation  of the youngest components of the Galactic thin disk 2-6 Gyr ago \citep{Montesinos2016}.

In the context of correlations with age and planet mass,  few works have been devoted to studies of DDP stars and fewer still have taken the DDP population as an ensemble. One of the few studies in that area was conducted by \citet{Maldonado2012}, where the authors looked for correlations between the debris disks properties, the presence of planets, and the metallicity of the central stars using a collection of 29 DDP stars with a well-determined $L_{dust}/L_{\star}$ ratio. The authors found that the presence of planets, rather than
the dust, is correlated with stellar metallicity. More recently, in  \citet[][hereafter CH19]{Chavero2019}, by using the mass of the debris disks ($M_{\rm d}$) instead of the fractional luminosity, the authors found a different result for the dust. In this study, the authors demonstrated that the $M_{\rm d}$ of the DDP stars is highly dependent on the stellar metallicity, which turns out to be smaller when the metallicity of the central star is lower. \\

The study presented here is an extension of the work in CH19, whereby we are studying the effects of Galactic events, such as thick and thin Galactic stellar disk formation, on the properties of the second generation of circumstellar disks (i.e., so-called debris disks). It is well known that metallic material is necessary to form planets \citep[e.g.,][]{Chambers2001}, but it has not been exhaustively studied. In particular, research is lacking with regard to the fact that when iron abundances are low
(i.e., in the initial epochs of the Galactic life when the thick-disk stars were formed at a [Fe/H]$<-0.2$), other
abundant elements (with condensation temperatures similar to that of iron) can replace the insufficient iron level -- and ultimately lead to planetary formation. These elements are the $\alpha$ elements that are known to have been relatively abundant by the time of the formation of the thick disk \citep[e.g.,][]{Bensby2005,Ishigaki2012}. The relative abundance of iron and alpha elements, [$\alpha$/Fe], changed over time, following an increase in SNIa explosions \citep{Adibekyan2012,Haywood2013,Bensby2014}. This
expected evolution on the alpha versus iron ratio has been extensively studied and, together with the stellar kinematics, it is now used to determine which Galactic population each star belongs to (halo, thick disk, or thin disk). Here, we study whether this alpha and/or iron evolution has indeed had an impact on the DDP population. Other aspects we analyze here concern the relation between the central star metallicity and the formation of DDs. Finally, we also explore whether there is a limit on the metallicity that qualifies stars as belonging to the DD, DDP, or CP systems by looking for the oldest thick-disk stars with very low metallicities.
In our analysis, and with the aim of reducing the parameters space, we focus only on solar-type main-sequence stars with spectral types between F5 up to K4  and stellar masses up to 1.4\,M$_{\odot}$.

The paper is organized as follows. In Sect.~\ref{sec:data}, we present the observational data used in this work and the process we used to to determine the Galactic population of each star (thin or thick disks). In Sect.~\ref{sec:diskmass}, we calculate the disk masses of the oldest DDP stars. In Sect.~\ref{sec:metallicity}, we
study the correlation between metallicity and the dust disk mass of the DDP and DD stars and we studied the
lithium evolution for the ensemble of our stars. Finally, we discuss our results and present our conclusions in Sect.~\ref{sec:conc}.

\section{Observational data: DDP, DD, and CP stars}\label{sec:data}

In this section, we present the data sample of stars that we use all throughout this work (see Table~\ref{t:table1}) and the process we used to determine which stars belong to the thin or the thick disks. We are not interested on the oldest and kinematically hottest population that is located in the halo, thus we only kept the stars belonging to the thin or the thick disks. To determine their membership, we employed two different approaches: the first based on the [$\alpha$/Fe] versus [Fe/H] analysis and the second based on the stellar kinematics provided by {\em Gaia}. 
Table~\ref{t:table1} presents the main properties of our sample. The values for all the stellar parameter have been obtained from the literature. $^a$ References of the stellar parameters (M$_\star$, Age, [Fe/H]); $^b$ References for the [Ti/H], [Ti/Fe], [O/H], [Mg/H],[Si/H] and $^c$ References for the A(Li). These correspond to:
Ad12 \citep{Adibekyan2012}; AG18  \citep{Aguilera-Gomez2018}; Be18 \citep{Bensby2018}; Ber18 \citep{Berger2018}; Bon16 \citep{Bonfanti2016}; Br16 \citep{Brewer2016}; Ca11 \citep{Casagrande2011}; CH19 \citep{Chavero2019}; DM15 \citep{DelMe2015};
DM17 \citep{DelMe2017}; DM19 \citep{DelMe2019}; Ed93 \citep{Edvardsson1993}; Ghe10 \citep{Ghezzi2010}; Hol09 \citep{Holmberg2009}; Lu17 \citep{Luck2017}; Pa13 \citep{Pace2013}; Ra12 \citep{Ramirez2012}; Ra14 \citep{Ramirez2014}; SilAg15 \citep{SilAg2015}; So05 \citep{Soubiran2005}; So18 \citep{Sosa18}; Ta20 \citep{Taut20}; Va05 \citep{ValentiFischer2005}.
For the few stars with no [Ti/H] values available, we assumed [Ti/Fe]=[$\alpha$/Fe] and tagged them with the * symbol; <O,Mg,Si/Fe> means the average
of [O/Fe], [Mg/Fe], and [Si/Fe]. Stars identified as high-$\alpha$ high-metal are tagged with a ** symbol.
The galactocentric velocity  values, $U,$ were taken   \citet{llorente2021}.
Exoplanet parameters data have been obtained from the Extrasolar Planets Encyclopaedia, with N$_p$ being the number of planets around the central star and  $\sum$ $M_{\rm pl}$ as the sum of the mass of all planets in the planetary system. We have taken the debris disk mass  (log($M_{\rm d}$)) in $M_{\rm Moon}$ units from \citet{Chen2014}, which does not include  the stars HD\,10700, HD\,20974, and HD\,40307; therefore, their disk masses were calculated within the scope of this work (see Sect.~\ref{sec:diskmass}).

\begin{sidewaystable*}

  \centering
\caption{Stellar sample studied in this work  }
\label{t:table1}
\scalebox{0.70}{
\begin{tabular}{lcccccccccccccccccc} 
\hline
\hline
Star         & Class &  Disk   & SpType    &   M$_\star$         &  Age                            & [Fe/H]             &  Ref.$^a$        & [Ti/H]    &  [Ti/Fe]    & <O,Mg,Si/Fe>  & Ref.$^b$ & U          & A(Li) & Ref.$^c$ & N$_p$ &  $\sum$ $M_{\rm pl}$   &   log($M_{\rm d} $)\\  
             &       &        &         &($\mathcal{M}_\odot$)   & (Gyr)                           & (dex)              &                  &           &             &               &          &   kms$^{-1}$&      &     &       &($M_{\rm Jup}$)         &($M_{\rm Moon}$)  \\   
\hline
HD 20759     & DD    &  Thin   & F5~V      &  1.25  $\pm$ 0.03   &  3.2$^{+1.4}_{-0.2}$            & -0.38  $\pm$ 0.10  &  Hol09,Ca11      &   -       &    0.15*    &  ---   &  Ca11       &  -42.00    &       &        & -    &  -         &   -5.51    \\
HD 101259    & DD    &  Thick  & G6/G8~V   &  1.53  $\pm$ 0.07   &  6.8$^{+4.8}_{-3.9}$            & -0.746 $\pm$ 0.014 &  Ho19,Va05,Ca11  &  -0.36    &    0.38     &  0.31  &  Va05,Be18  &  35.88     & 0.80  &  Be18  & -    &  -         &   -4.32    \\
HD 110897    & DD    &  Thick  & F9~V      &  0.96 $\pm$ 0.03    &  5.5 $\pm$ 3.8                  & -0.59  $\pm$ 0.05  &  CH19            &  -0.32    &    0.27     &  0.26  &  Br16       &  -42.00       & 1.94  &  Lu17   & -    &  -         &   -1.95    \\
HD 154577    & DD    &  Thin   & K1~V      &  0.642 $\pm$  0.014 &  7.14 $\pm$ 4.27                & -0.650  $\pm$ 0.018 &  DM19,Ho19       &  -0.55    &    0.14     &  0.08  &  Va05,Ad12  &  -10.05    & 0.51  &  DM15   & -    &  -         &   -4.20    \\
HD 158633    & DD    &  Thin   & K0~V      &  0.94 $\pm$ 0.13    &  7.8$^{+5.2}_{-6.4}$            & -0.45  $\pm$ 0.12  &  Va05,Ta20       &  -0.34    &    0.11     &  0.15  &  Br16       &  1.41      & 0.11  &  CH19   & -    &  -         &   -2.30    \\
HD 213941    & DD    &  Thick  & G8        &  0.799 $\pm$ 0.009  &  12.811 $\pm$ 0.536             & -0.46  $\pm$ 0.01  &  DM19            &  -0.18    &    0.28     &  0.24  &  Ad12       &  -5.98     & 0.34  &  CH19   & -    &  -         &   -5.96    \\
HD 6434      & CP    &  Thick  & G2/3~V    &  0.82 $\pm$ 0.02    &  9.55 $\pm$ 2.95                & -0.61  $\pm$ 0.04  &  Pa13,AG18       &  -        &    0.33*    &  0.37  &  So18,Be18  &  89.58     & 0.84  &  AG18   & 1    &  0.39      &   -        \\
HD 37124     & CP    &  Tran   & G4~IV/V   &  0.82 $\pm$ 0.02    &  11.10 $\pm$ 1.70               & -0.43  $\pm$ 0.01  &  Bon16,So18      &  -0.16    &    0.27     &  0.34  &  Br16       &  32.14     & 0.49  &  AG18   & 3    &  2.02      &   -        \\
HD 39194     & CP    &  Thick  & K0~V      &  0.709 $\pm$ 0.014  &  10.441 $\pm$ 2.540             & -0.610 $\pm$ 0.012 &  DM19,Ho19       &  -0.24    &    0.37     &  0.24  &  Ad12       &            & ---   &  ----   & 3    &  0.05      &   -        \\
HD 102365    & CP    &  Thin   & G2~V      &  0.86 $\pm$ 0.03    &  8.5  $\pm$ 2.8                 & -0.35  $\pm$ 0.03  &  CH19            &  -0.18    &    0.11     &  0.15  &  Br16       &  -60.14    & 0.52  &  CH19   & 1    &  0.05      &   -        \\
HD 114729    & CP    &  Thick  & G0~V      &  0.92 $\pm$ 0.01    &  10.8 $\pm$ 0.5                 & -0.33  $\pm$ 0.06  &  CH19            &  -0.08    &    0.20     &  0.17  &  Va05,Ad12  &  18.62     & 1.88  &  CH19   & 1    &  0.84      &   -        \\
HD 136352    & CP    &  Thick  & G3/5~V    &  0.87 $\pm$ 0.02    &  9.6  $\pm$ 1.8                 & -0.28  $\pm$ 0.07  &  CH19            &  -0.12    &    0.22     &  0.17  &  Br16       &  -119.25   & 0.86  &  CH19   & 3    &  0.08      &   -        \\
HD 154857    & CP    &  Thin   & G5~V      &  1.21 $\pm$ 0.06    &  4.9$^{+0.7}_{-0.6}$           & -0.26  $\pm$ 0.01  &  Hol09,So18      &  -0.15    &    0.11     &  0.07  &  Va05       &  20.43     & 1.64  &  CH19   & 1    &  5.82      &   -        \\
HD 155358    & CP    &  Thin   & G0        &  0.89 $\pm$ 0.02    & 10.37 $\pm$ 1.23                & -0.65  $\pm$ 0.04  &  Pa13,AG18       &  -0.52    &    0.13     &  0.13  &  Ed93       &  28.86     & 2.14  &  CH19   & 2    &  1.67      &   -        \\
HIP 94931 &CP&  Thick  & K0~V      &  0.739 $\pm$ 0.009  & 11.54 $\pm$ 0.99                & -0.37  $\pm$ 0.09  &  SilAg15         &  -0.23    &    0.29     &  0.19  &  Br16       &            & 0.34  &  Ber18   & 5    &  0.3       &   -        \\
HD 189567    & CP    &  Thick  & G2~V      &  0.90 $\pm$ 0.06    & 11.0 $\pm$ 0.5                  & -0.24  $\pm$ 0.04  &  CH19            & - 0.141   &    0.13     &  0.081 &  Ad12,Va05  &  -70.89    & 0.86  &  CH19   & 2    &  0.05      &            \\
HD 1461      & DDP   &  Thin   & G0~V      &  1.05 $\pm$ 0.03    &  4.3 $\pm$ 2.5                  &  0.18  $\pm$ 0.04  &  CH19            &  0.15     &    0.03     &  -0.04 &  Br16       &  -31.74    & 0.60  &  CH19   & 2    &  0.04      &   -1.58    \\
HD 10647     & DDP   &  Thin   & F8~V      &  1.10 $\pm$ 0.03    &  1.6 $\pm$ 1.3                  & -0.05  $\pm$ 0.04  &  CH19            &  -0.11    &   -0.11     &  -0.06 &  Va05       &  -1.12     & 2.74  &  CH19   & 1    &  0.93      &   0.04     \\
HD 10700     & DDP   &  Thick  & G8~V      &  0.77 $\pm$ 0.02    &  8.2 $\pm$ 3.2                  & -0.52  $\pm$ 0.07  &  CH19            &  -0.32    &    0.21     &  0.20  &  Ad12       &  18.72     & 0.42  &  CH19   & 2    &  0.025     &   -4.64    \\
HD 20794     & DDP   &  Thick  & G8~V      &  0.81 $\pm$ 0.03    & 11.4 $\pm$ 0.3                  & -0.41  $\pm$ 0.03  &  CH19            &  -0.12    &    0.28     &  0.26  &  Ad12       &  -78.4     & 0.58  &  Ghe10  & 3    &  0.03      &   -5.16    \\
HD 22049     & DDP   &  Thin   & K2~V      &  0.79 $\pm$ 0.02    &  3.8 $\pm$ 3.6                  & -0.15  $\pm$ 0.06  &  CH19            &  -0.03    &    0.12**   &  0.11  &  Br16       &  -3.66     & 0.47  &  CH19   & 1    &  0.65      &   -3.66    \\
HD 33636     & DDP   &  Thin   & G0        &  1.01 $\pm$ 0.02    &  1.3 $\pm$ 1.3                  & -0.10  $\pm$ 0.04  &  CH19            &  -0.03    &    0.07     &  0.05  &  Br16       &  -0.17     & 2.4   &  CH19   & 1    &  9.24      &   -1.52    \\
HD 38858     & DDP   &  Thin   & G4~V      &  0.89 $\pm$ 0.02    &  7.3 $\pm$ 1.4                  & -0.22  $\pm$ 0.04  &  CH19            &  -0.21    &    0.01     &  0.06  &  Br16       &  -17.41    & 1.57  &  Lu17   & 1    &  0.10      &  -1.35     \\
HD 39091     & DDP   &  Thin   & G0~V      &  1.10 $\pm$ 0.03    &  3.0 $\pm$ 1.8                  &  0.11  $\pm$ 0.06  &  CH19            &  0.02     &   -0.09     &  -0.01 &  Va05,Ad12  &  -82.98    & 2.25  &  CH19   & 1    & 10.30      &   -2.84    \\
HD 40307     & DDP   &  Thick  & K3~V      &  0.70 $\pm$ 0.01    &  6.0 $\pm$ 4.1                  & -0.36  $\pm$ 0.03  &  CH19            &  -0.09    &    0.27     &  0.13  &  Ad12       &  3.02      & 0.22  &  Ghe10  & 6    &  0.104     &   -4.68    \\
HD 40979     & DDP   &  Thin   & F8        &  1.21 $\pm$ 0.02    &  0.8 $\pm$ 0.6                  &  0.21  $\pm$ 0.06  &  CH19            &  0.24     &    0.03     &  0.04  &  Br16       &  -36.62    & 2.91  &  CH19   & 1    &  4.01      &   -2.86    \\
HD 45184     & DDP   &  Thin   & G2~V      &  1.00 $\pm$ 0.01    &  4.4 $\pm$ 2.7                  &  0.04  $\pm$ 0.03  &  CH19            &  0.05     &    0.01     &  0.00  &  Br16       &  10.04     & 2.01  &  CH19   & 1    &  0.04      &   -0.39    \\
HD 50499     & DDP   &  Thin   & G1~V      &  1.22 $\pm$ 0.03    &  3.8 $\pm$ 0.5                  &  0.22  $\pm$ 0.04  &  CH19            &  0.28     &    0.06     &  0.09  &  Br16       &  -32.55    & 2.62  &  CH19   & 1    &  1.71      &   -3.49    \\
HD 50554     & DDP   &  Thin   & F8~V      &  1.10 $\pm$ 0.03    &  2.1 $\pm$ 1.6                  &  0.05  $\pm$ 0.06  &  CH19            &  0.03     &   -0.02     &  -0.02 &  Br16       &  3.83      & 2.42  &  CH19   & 1    &  5.16      &   -1.64    \\
HD 52265     & DDP   &  Thin   & G0~V      &  1.20 $\pm$ 0.03    &  3.9 $\pm$ 1.0                  &  0.21  $\pm$ 0.04  &  CH19            &  0.19     &   -0.02     &  -0.01 &  Br16       &  -53.12    & 2.65  &  CH19   & 2    &  1.40      &   -1.85    \\
HD 82943     & DDP   &  Thin   & G0        &  1.20 $\pm$ 0.02    &  1.0 $\pm$ 1.0                  &  0.28  $\pm$ 0.03  &  CH19            &  0.22     &   -0.06     &  -0.04 &  Br16       &  22.9      & 2.47  &  Ghe10  & 1    &  0.29      &   -1.17    \\
HD 73526     & DDP   &  Thin   & G6~V      &  1.09 $\pm$ 0.05    &  5.0 $\pm$ 2.8                  &  0.27  $\pm$ 0.06  &  CH19            &  0.27     &    0.00     &  0.04  &  Br16       &  -78.91    & 0.59  &  CH19   & 2    &  4.50      &   -2.95    \\
HD 108874    & DDP   &  Thin   & G5        &  0.81 $\pm$ 0.05    &  6.2 $\pm$ 2.5                  &  0.20  $\pm$ 0.06  &  CH19            &  0.15     &   -0.05     &  -0.04 &  Br16       &  44.91     & 0.58  &  CH19   & 2    &  2.38      &   -1.96    \\
HD 113337    & DDP   &  Thin   & F6~V      &  1.40 $\pm$ 0.04    &  1.5 $\pm$ 0.9                  &  0.13  $\pm$ 0.04  &  CH19            &  -        &   -0.04*    &   ---  &  Ca11       &  -21.83    & ---   &  ----   & 2    & 10.3       &   -0.72    \\
HD 115617    & DDP   &  Thin   & G5~V      &  0.94 $\pm$ 0.04    &  3.5 $\pm$ 3.0                  &  0.02  $\pm$ 0.03  &  CH19            &  -0.01    &   -0.03     &  -0.03 &  Br16       &  -23.59    & 0.22  &  Ghe10  & 3    &  0.135     &   -2.24    \\
HD 117176    & DDP   &  Thin   & G5~V      &  1.06 $\pm$ 0.03    &  8.1 $\pm$ 0.4                  & -0.06  $\pm$ 0.03  &  CH19            &  -0.04    &    0.02     &  0.01  &  Br16       &  12.92     & 1.85  &  Ra12   & 1    &  6.60      &   -1.64    \\
HD 128311    & DDP   &  Thin   & K0        &  0.81 $\pm$ 0.04    &  1.0 $\pm$ 0.5                  &  0.01  $\pm$ 0.03  &  CH19            &  0.14     &    0.13**   &  0.10  &  Br16       &  16.56     & 0.13  &  CH19   & 2    &  6.37      &   -3.91    \\
HD 130322    & DDP   &  Thin   & K0~V      &  0.90 $\pm$ 0.03    &  4.1 $\pm$ 3.6                  &  0.01  $\pm$ 0.04  &  CH19            &  0.04     &    0.03     &  0.02  &  Br16       &  -9.72     & 0.38  &  CH19   & 1    &  1.05      &   -3.10    \\
HD 150706    & DDP   &  Thin   & G0        &  1.03 $\pm$ 0.02    &  0.3 $\pm$ 0.3                  & -0.01  $\pm$ 0.05  &  CH19            &  0.01     &    0.02     &  -0.02 &  Br16       &  19.71     & 2.52  &  CH19   & 1    &  2.71      &            \\
HD 178911 B  & DDP   &  Thin   & G5~V      &  1.02 $\pm$ 0.03    &  4.2 $\pm$ 1.8                  &  0.27  $\pm$ 0.04  &  CH19            &  0.20     &   -0.07     &  -0.08 &  Br16       &  -62.64    & 0.47  &  CH19   & 1    &  6.29      &   -        \\
HD 187085    & DDP   &  Thin   & G0~V      &  1.24 $\pm$ 0.03    &  1.2 $\pm$ 0.9                  &  0.21  $\pm$ 0.06  &  CH19            &  0.08     &   -0.13     &  -0.21 &  Va05,Ra14  &  11.11     & 2.66  &  CH19   & 1    &  0.75      &   -        \\
HD 192263    & DDP   &  Thin   & K0        &  0.80 $\pm$ 0.05    &  3.4 $\pm$ 3.4                  & -0.06  $\pm$ 0.05  &  CH19            &  0.06     &    0.12**   &  0.11  &  Br16       &  -16.09    & 0.02  &  CH19   & 1    &  0.73      &   -3.46    \\
HD 210277    & DDP   &  Thin   & G0        &  0.95 $\pm$ 0.02    &  5.0 $\pm$ 3.3                  &  0.18  $\pm$ 0.04  &  CH19            &  0.20     &    0.02     &  0.03  &  Br16       &  4.08      & 0.42  &  CH19   & 1    &  1.23      &   -2.50    \\
HD 215152    & DDP   &  Thin   & K0        &  0.76 $\pm$ 0.07    &  5.2 $\pm$ 4.0                  & -0.08  $\pm$ 0.09  &  CH19            &  0.10     &    1.82**   &  0.03  &  Ad12       &  22.82     & ---   &  ----   & 4    &  0.03      &   -4.39    \\
HD 216435    & DDP   &  Thin   & G3~IV     &  1.30 $\pm$ 0.03    &  3.4 $\pm$ 0.5                  &  0.20  $\pm$ 0.07  &  CH19            &  0.24     &    0.04     &  0.02  &  Va05,Ad12  &  -27.42    & 2.65  &  CH19   & 1    &  1.26      &   -1.96    \\
HD 224693   &  DDP   &  Thin   & G2~V       &  1.31 $\pm$ 0.09    &  3.4 $\pm$ 3.4                  & 0.28   $\pm$ 0.07  &  CH19            &  0.25     &   -0.03     &  -0.02 &  Br16       &  -62.50    & 2.03  &  CH19   & 1    &  0.71      &   -        \\
\hline 
\end{tabular}}
\tablefoot{ 
*: [Ti/Fe]=[$\alpha$/Fe]. 
**: Stars identified as HAHM 
}
\end{sidewaystable*}


\subsection{Abundances of [$\alpha$/Fe]  versus} [Fe/H] \label{subsec:chem}

\begin{figure*}[!]
\centering
  \subfigure[\hbox{ [Ti/Fe] versus [Fe/H]}]{\includegraphics{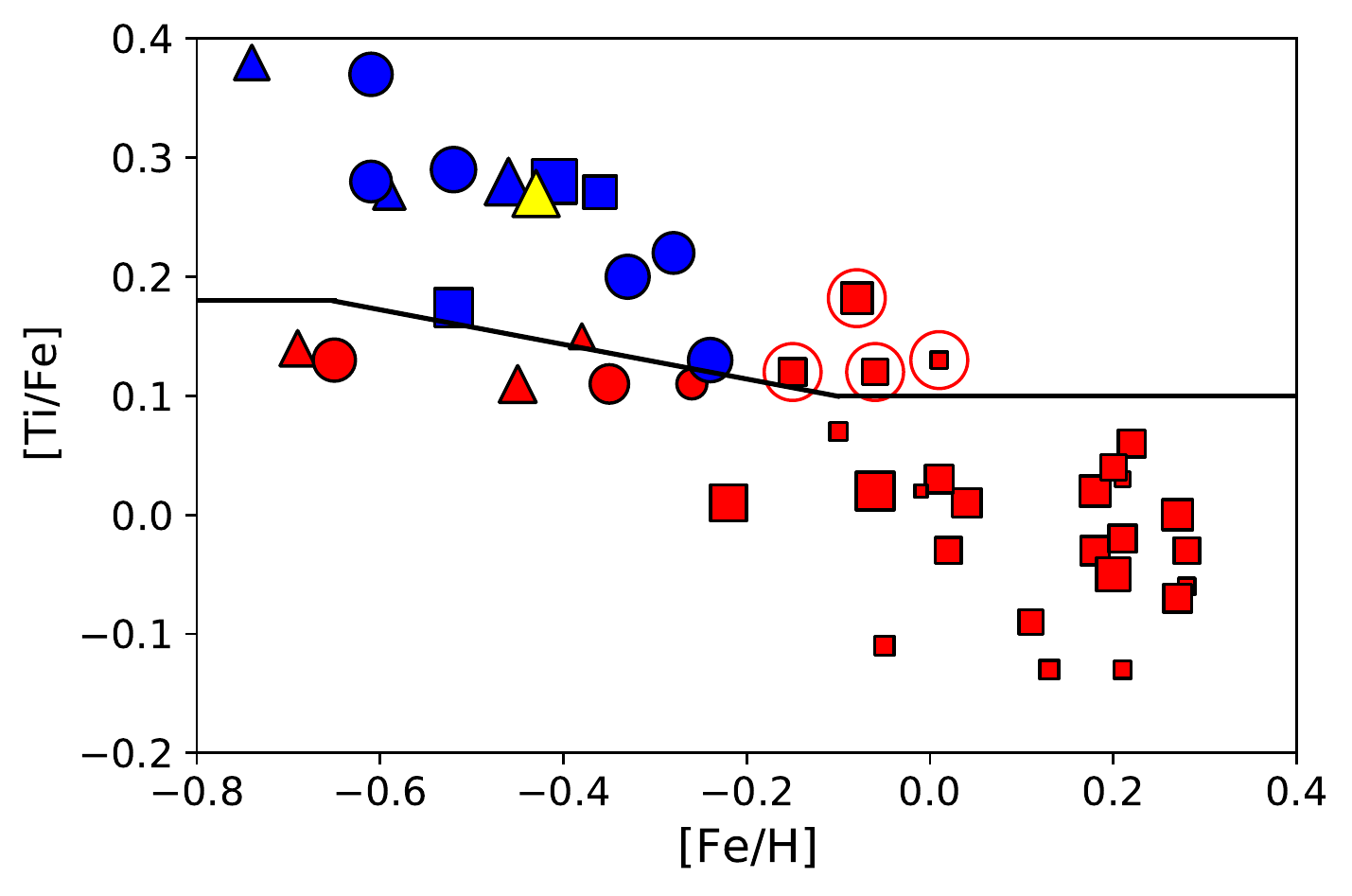}}
  \hfill
    \subfigure[\hbox{  Average between [O/Fe], [Si/Fe] and [Mg/Fe] versus [Fe/H]}]{\includegraphics{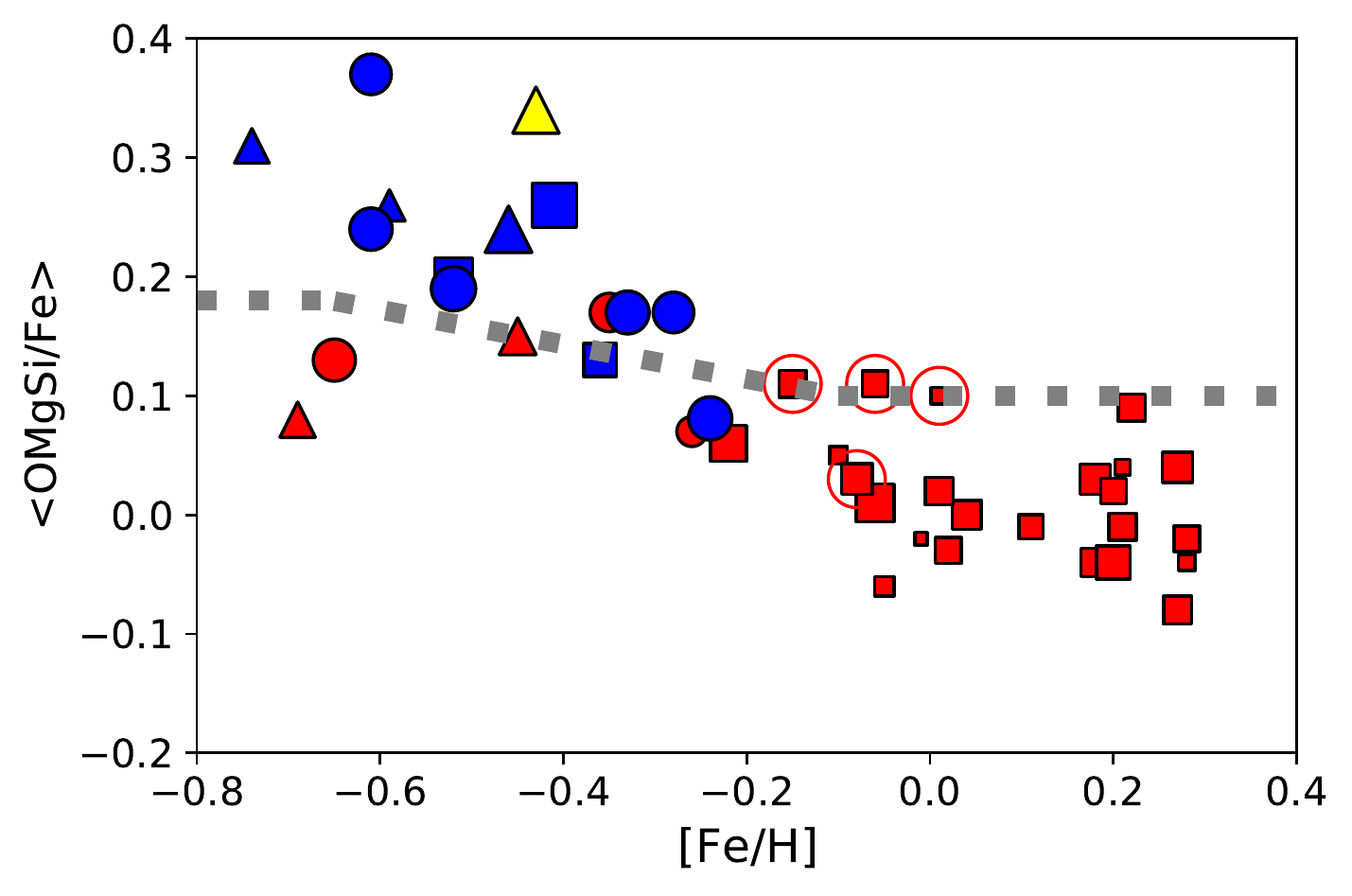}}
  \hfill
\caption{ [$\alpha$/Fe] versus [Fe/H].
 Panel (a): Distribution of abundances of [Ti/Fe] versus [Fe/H] of stars in Table~\ref{t:table1}. 
Symbols indicate the star's class: DDP (squares), DD (triangles), CP (circles). The black 
line separates the stars as belonging to the thin (below) or thick (above) disks following the definition presented by \citet{Bensby2014}. Colors show stars that fulfil both the chemistry-based and the kinematic criteria to belong to the thin (red) and/or the thick (blue) disk populations (see Sects.~\ref{subsec:chem} and ~\ref{subsec:kinem}). In yellow we show the stars that are in a “transitory stage” between the two disk populations according to its kinematics. Stars surrounded by a red circle indicate they belong to the HAHM population, as defined by \cite{Adibekyan2012}. Finally, the size of the symbols qualitatively indicate the stellar ages, being the smallest the youngest.
Panel (b): Same details as in panel (a) but considering the average between [O/Fe], [Si/Fe], and [Mg/Fe] as a function of [Fe/H]. The grey dashed line shows the limit between the thin- and thick-disk stars   in the [$\alpha$/Fe]  versus [Fe/H] space.
 }
 \label{fig1}
\end{figure*}

In Figure~8 of \citet{Adibekyan2012}, the authors presented a collection of stars with well determined abundances of many $\alpha$-elements, as well as a good determinations of the [Fe/H]. Among the $\alpha$--elements they analyzed, the sample of [Ti/H] is the most clear and complete. Thanks to these proper determinations, Ti has already been used to study the relation of  $\alpha$--elements with stellar mass \citep[e.g.,][]{Bensby2014,Campante2015}. Thus, we also chose to use Ti as a proxy for the $\alpha$--elements in our study. All [Ti/H] values we used are included in Table~\ref{t:table1}, together with their references. In Table~\ref{t:table1}, we also show the [Fe/H] and [Ti/Fe]. We display the distribution of [Ti/Fe] as a function of [Fe/H] in Fig.~\ref{fig1}. In this figure, we also draw a black solid line that splits (as a first approximation) stars belonging to the thin disk, shown below, from stars in the thick disk, shown above \citep[see][]{Bensby2014}. In Sect.~\ref{sec:conc}, we further discuss  the finding of four new stars in the category of high-alpha high-metallicity stars (HAHM), first identified by \citet{Adibekyan2012}, which are marked with red circles in this Fig.~\ref{fig1}. These four stars group in a region that seems to not belong to the thin or the thick disk; that is to say, they may have a different origin (see Sect.~\ref{sec:conc} for a further discussion). In Fig.~\ref{fig1}, we also show qualitatively the stellar ages (the symbol sizes grow with age) and the results are in agreement with an expected decay in the iron abundances with stellar age.

\subsection{     [O, Mg, Si] versus [Fe/H]}    

In the previous section, we describe the use of titanium as a proxy for the alpha elements, following works by other authors. However, some authors have reported several peculiarities in the formation of this element at specific metallicity ranges that may affect this decision \citep[see the review paper by][]{McWilliam1997}. To make our conclusions more robust, in this section, we use the average abundance of oxygen, magnesium, and silicon as proxies of the alpha elements. We compare these results with those we obtained when using titanium. We use the average value to reduce the dispersion on the alpha values as suggested by \citet{Lambert1987}.
Our main objective here is to demonstrate that Ti shows a similar metallicity distribution as <O, Mg, Si>. We show our results in Figure~\ref{fig1} (panels a and  b). As a result (excluding a few particular cases), the [Ti/Fe] versus [Fe/H] distribution (Figure~\ref{fig1}a, left panel) follows the same trend as  [<O,Mg,Si>/Fe] versus [Fe/H] one (Figure~\ref{fig1}b, right panel). The more noticeable difference between both distributions appears for the peculiar stars HAHM belonging to the thin disk (mentioned in Sect. 2.1). These stars (surrounded by a red circle) appear much better defined by Ti than by the O, Mg, and Si
elements.
In light of these results, we conclude that both Ti and the <O, Mg, Si> are good tracers of the alpha elements, but also that Ti allows us to better discriminate between thin- and thick-disk stars (symbols with the same colors clump better). Also, we state that our conclusions remain the same if changing from one to the other group of elements. The median of the  [O/Fe], [Mg/Fe], and [Si/Fe]  abundances are presented in Table~\ref{t:table1} as <O, Mg, Si /Fe>, with the respective references.

\subsection{Stellar kinematics}\label{subsec:kinem}  

\begin{figure}
     \centering
   \includegraphics[width=7.5cm]{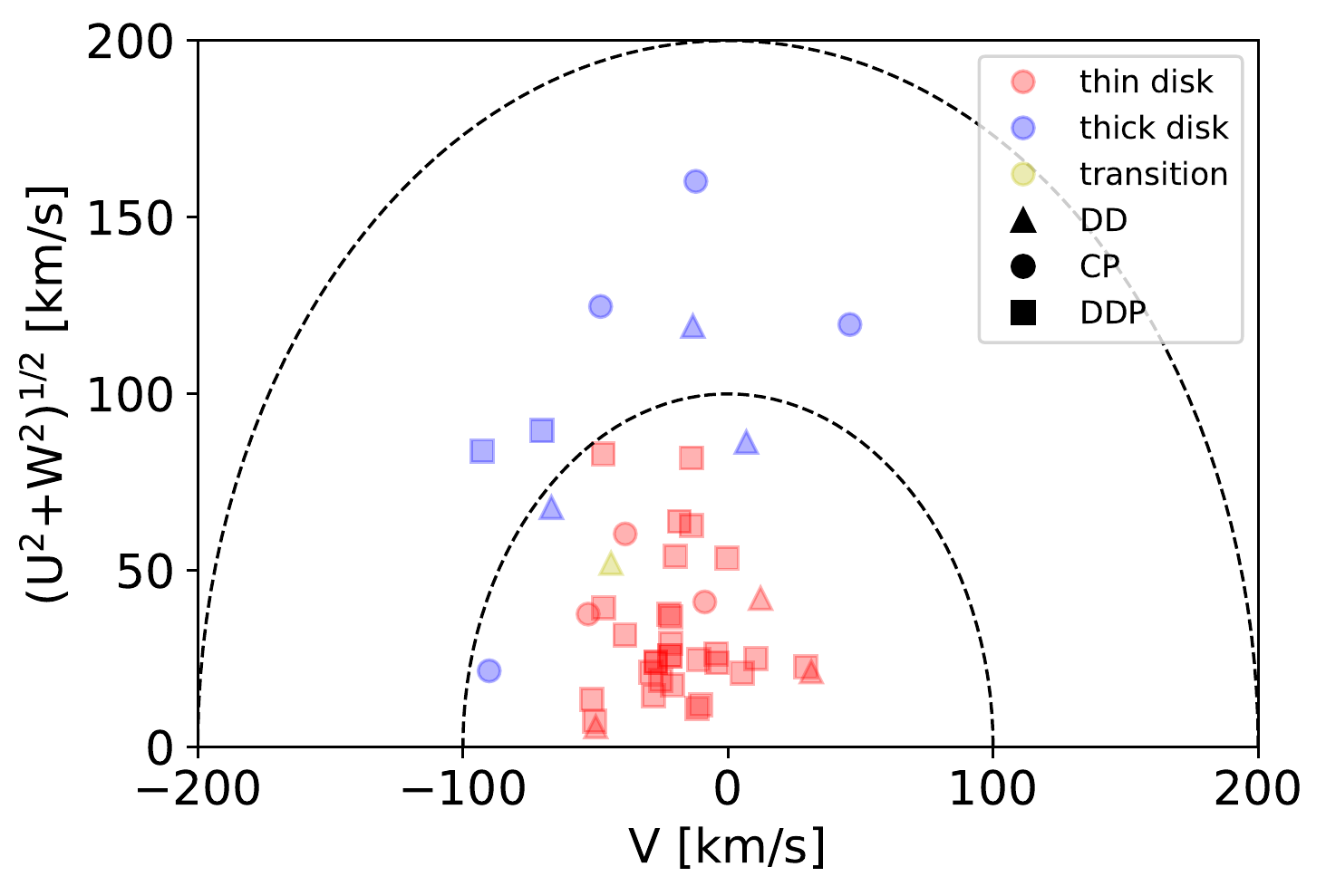}\par
   \caption{Toomre diagram of the stars presented in Table~\ref{t:table1}. Symbols and colors are
the same as in Fig.~\ref{fig1}. The Galactic space velocities (U, V, W) are
relative to the local standard of rest. Dashed lines represent curves of
constant peculiar velocities of 100 and 200\,kms$^{-1}$. 
}
   \label{fig2}
    \end{figure}

As a second method used to differentiate between thin- and thick-disk stars, we followed the approach presented by \citet{Bensby2014}, detailed in their Appendix~A. This method is based on a detailed analysis of the stellar kinematics, which is now more reliable thanks to the new high-precision data produced by the {\em Gaia} satellite \citep{Gaia2018}.  We used the U, V, and W galactocentric velocities relative to the local standard of rest  that we obtained, published in \citet{RocaFabrega21}, to study the probability of belonging to the thin-disk, thick-disk, halo, or transitional phases, as in \citet{Bensby2014}. We show the results of this analysis in Fig.~\ref{fig2} and there we can clearly see the presence of a thin-disk population of stars that have velocities lower than 50\,kms$^{-1}$ and a thick disk that is made up of stars with velocities between 70 and 200\,kms$^{-1}$. Also, stars with velocities between 50 and 70\,kms$^{-1}$ are the ones considered to be transitory members between the thin and thick disks; thus, they require additional information on their chemistry to be classified in one or another population (see Fig.~\ref{fig1}). The need for using the two criteria together to better define the thin- and thick-disk populations was also discussed in many previous papers \citep[see, e.g.,][]{Navarro2011}. Finally, we adopt here the following criteria: if a star fulfill both, the chemical and kinematical conditions of membership to a given Galactic disk population, the star is classified accordingly. Stars with a disagreement between methods are  analyzed in depth. In general, these outliers are considered thin-disk stars if they have high metallicity, as they can be kinematically heated up by intrinsic disk perturbations; they are considered thick-disk stars if they are extremely metal poor, as (although it is unusual) thick-disk stars with low velocity dispersion can also fit within the thick-disk velocity distribution. 
We are aware that these last conditions can appear arbitrary, but they have no strong impact on our results, as the vast majority of objects of Table~\ref{t:table1} fulfill both conditions well. We note that only one member was found in the transitory kinematical case (see yellow triangle in Fig.~\ref{fig1}).

\section{Debris disk masses}\label{sec:diskmass}

In general, due to their very low metallicities, old and very old DD and DDP stars are expected to host low-mass debris disks and, as a consequence, low-mass planets \citep[e.g.,][]{Courcol2016,Sousa2019}.
With the goal of maintaining uniformity and homogeneity, in this work we only analyze the debris disk masses ($M_{\rm d}$) obtained from unresolved disks using the methodology presented by \citet{Chen2014}. We also applied this method to the
few objects with resolved debris disks we found in the literature. Regarding the objects detailed in \citet{Chen2014}, as the debris
disk mass, we only considered  the second and the largest $M_{\rm d2}$ values if they were based on a two-zone
model -- and  $M_{\rm d}$, otherwise. In this section we focus on three important objects that are not included in the \citet{Chen2014} catalog: HD\,10700, HD\,20794, and HD\,40307, (hereafter, the "trio"). These three
stars show extreme low metallicity and, thus, this trio could be the oldest DDP stars ever found. In the following, we present a determination of their debris disk mass and of the total planetary masses and number of planets, however, we do not go into the detail about the planets' orbital and mass structures.\\

{\bf HD\,20794:} \citet{Kennedy2014} carried out a very detailed study of the resolved debris disk of HD\,20794 G8~V star using Hershel and JCMT sub-mm data. Their results show that its debris disk is only marginally 
resolved and much fainter than the one around HD\,38858 G4~V star, which is also a DDP star. We hypothesise that this difference is intrinsic to variations on the properties of the protostellar clump where they 
were born. In fact, as we demonstrate in Sect.~\ref{sec:data},  HD\,20794 belongs to the Galactic thick disk, while  HD\,38858 is a member of the Galactic thin disk (see Table~\ref{t:table1}). This hypothesis and 
the results agree  with the age estimations for HD\,20794 well, namely, 11.3 $\pm$ 0.03\,Gyr, according to our work in CH19 and the one by \citet{Holmberg2009}; and for HD\,38858 that result is 7.3 $\pm$ 1.4\,Gyr (CH19). 
Going back to  HD\,20794 disk properties, \citet{Kennedy2014} proposed that the observations can be aptly explained with models of a 24 AU narrow debris ring, with a width of 5 AU. These results suggest that HD\,20794's disk
is actually one of the faintest debris disks known thus far with robust photometric measurements. We also determined that this thick-disk star is currently the oldest known DDP star.\\

{\bf HD\,10700 -- Tau\,Cet:} \citet{Greaves2004a} carried out the first study of the resolved disk of HD\,10700, that  is, Tau\,Cet  (a G8~V), using the sub-millimetre SCUBA camera. In contrast to the star HD\,20975 presented 
in the previous paragraph, the disk of Tau\,Cet appears to be extended. The models presented by \citet{Greaves2004a} suggest that most of the dust lies in a ring with a maximum radius of 55\,AU. However, further observations 
of Tau\,Cet now with ALMA at 1.3\,mm  \citep{MacGregor2016} at a much higher resolution \citep[see e.g.,  ][]{Lawler2014} revealed a patchy image with a complex radial structure. The new images showed that there is
an inner belt with its edge at $\sim$\,6\,AU accompanied by an external broad disk much wider than the Kuiper belt at our Solar System.\\

{\bf HD\,40307:} There is no detailed information about the debris disk of HD\,40307. We consider the debris disk of this star as “unresolved.”\\

To estimate the $M_{\rm d}$ of disk debris, it is necessary to know the $L_{IR}/L_*$ ratio and the disk radius. This is a problem for at least two of the three cases presented above, as it is not easy to obtain realistic radii
from the complex resolved disks. In light of these difficulties (and for the sake of homogeneity),  for the radii of the debris disks we decided to take the measurements obtained from the blackbody fitting, using a methodology
similar to the one used in the reference DD catalog  \citep{Chen2014}. We used  IR at 70\,$\mu$m photometry obtained with the Hershel telescope,  as done by  \citet{Sibthorpe2018} for HD\,20794 and HD\,10700 and  
\citet{Marshall2014} for HD\,40307. Details on the calculations we carried out to obtain the $M_{\rm d}$ values for these three stars can be found in Sect.~\ref{app:1} and the result is shown in Table~\ref{t:table2}.
Debris disk masses for these three stars when measured at 70\,$\mu$m are very small when compared with stars at a similar evolutionary state. It is important to recall that these stars are considered thick-disk stars, while the the other DDP stars we can use to compare the debris disk mass are always thin-disk stars, namely, they are born in molecular clouds with potentially different chemistry properties. Therefore, the low mass of the debris disks
around these three thick-disk DDP stars should not be surprising if it is correlated with metallicity, but this is a topic discussed further in the remainder of the paper. Finally, it is important to notice that although age estimations 
have still large uncertainties, these three stars seem to be older than the other DDP examples (see Table~\ref{t:table3}); this is a result that agrees well with their photospheric lithium abundances, which are small as a result of 
lithium depletion during the main sequence \citep[e.g.,][]{RocaFabrega21}.\\

Now, to get a better idea how small the masses of these debris disks really are, we compare them with the size of the mass of the Solar System Kuiper Belt (KB). The most recent values of the KB mass obtained from studies of the Solar
System dynamics are M$_{KB}$=  4.8\,$M_{Moon}$ \citep{DiRuscio2020} and  1.6\,$M_{Moon}$ \citep{Pitjeva2018}. These values are much larger than the $M_{\rm d}$ values for those of the debris disks in the stars of the trio,
by  up to five orders of magnitude (see Table~\ref{t:table3}). In addition, this star trio  formed (in agreement with their low disk masses)  equally low-mass planets. In fact, the sums of their planetary content in Jupiter
mass are: 0.034\,$M_{Jup}$ for HD\,20974, 0.025\,$M_{Jup}$ for HD\,10700, and 0.114\,$M_{Jup}$ for HD\,40307, which is the most metallic of this group.\\
Finally, it is important to mention that the Tau-Ceti is peculiar with regard to its debris disk mass if measured at 850\,$\mu$m. At this wavelength, \citet{Greaves2004a} showed that the disk presents a mass excess of 1.2 Earth masses.
If we consider that these old stars have normal old decaying disks, this is an unnaturally massive disk. Currently, the question of whether this massive belt is a remnant of the primordial debris disk or whether it has recently regenerated remains a puzzling
enigma. It is also an enigma of how frequent can these "regeneration events" occur and if we should expect to find these massive disks in other DDP stars.

\begin{center}
\begin{table}
\caption{Parameters of the trio of DDP stars detailed in Sect. 3, including the spectral type, their classification as thin- or thick-disk stars, metallicity, age, and debris disk mass as computed in this work, along with the number of known planets and total mass in planets.}
\label{t:table2}

\scalebox{0.75}{
\begin{tabular}{lccccccc}
\hline
\hline
Star & SpType    & Disk &  [Fe/H]  & Age   & $M_{\rm d}$       & Np &       $\sum$ $M_{\rm pl}$   \\
      &           &      & (dex)     & (Gyr) & ($M_{\rm (Moon)}$)   &  &$M_{\rm (Jup)}$         \\\hline \\

HD20794     & G8~V    &  Thick   & -0.41       &11.40 $\pm$ 0.3    & 7.09$\,\times\,10^{-6}$  &  3          &  0.034             \\
HD10700     & G8~V    &  Thick   & -0.52      & 8.20 $\pm$ 3.20    & 2.30$\,\times\,10^{-5}$  &  2          & 0.025               \\  
HD40307     & K3~V    &  Thick   &  -0.36      & 6.00 $\pm$ 4.10    & 2.10$\,\times\,10^{-5}$  &   6         &    0.104          \\  
\\
\hline 
\end{tabular}}
\end{table}
\end{center}

\section{Metallicity in DDP and DD stars}\label{sec:metallicity}

In this paper, we focus on the study of dwarf stars with spectral types between F5 and K4, covering a wide range of stellar ages. It has long been  established that a correlation exists between age and the iron and $\alpha$-element abundances \citep{Soubiran2008,Nissen2015}. This is not surprising as most of the element abundances reflect those of the interstellar medium (ISM) where the star was born. In this context, as the ISM properties follow the Galactic evolution, stellar abundances should also reflect the major events occurred in the Galaxy’s evolution \citep{RocaFabrega21}. Many previous works have shown that it also exists a correlation between the mass in planets and the metallicity of the host star (e.g., CH19). This result suggests that the total mass and properties of planets around stars from the thin-, thick-disk, and halo Galactic populations, are expected to be different. In this section, we  study possible correlations between the metallicity and the stellar and accumulated planetary masses further in a sample of DD and DDP thin- and thick-disk stars. Following our previous results, we expect to find positive gradients in the mass of DDP and/or DD stars and in their total planetary mass as function of [Fe/H]. \\

\subsection{Metallicity gradient of DDP stars}\label{subsec:metalDDP}

\begin{figure}
     \centering
   \includegraphics[width=7.5cm]{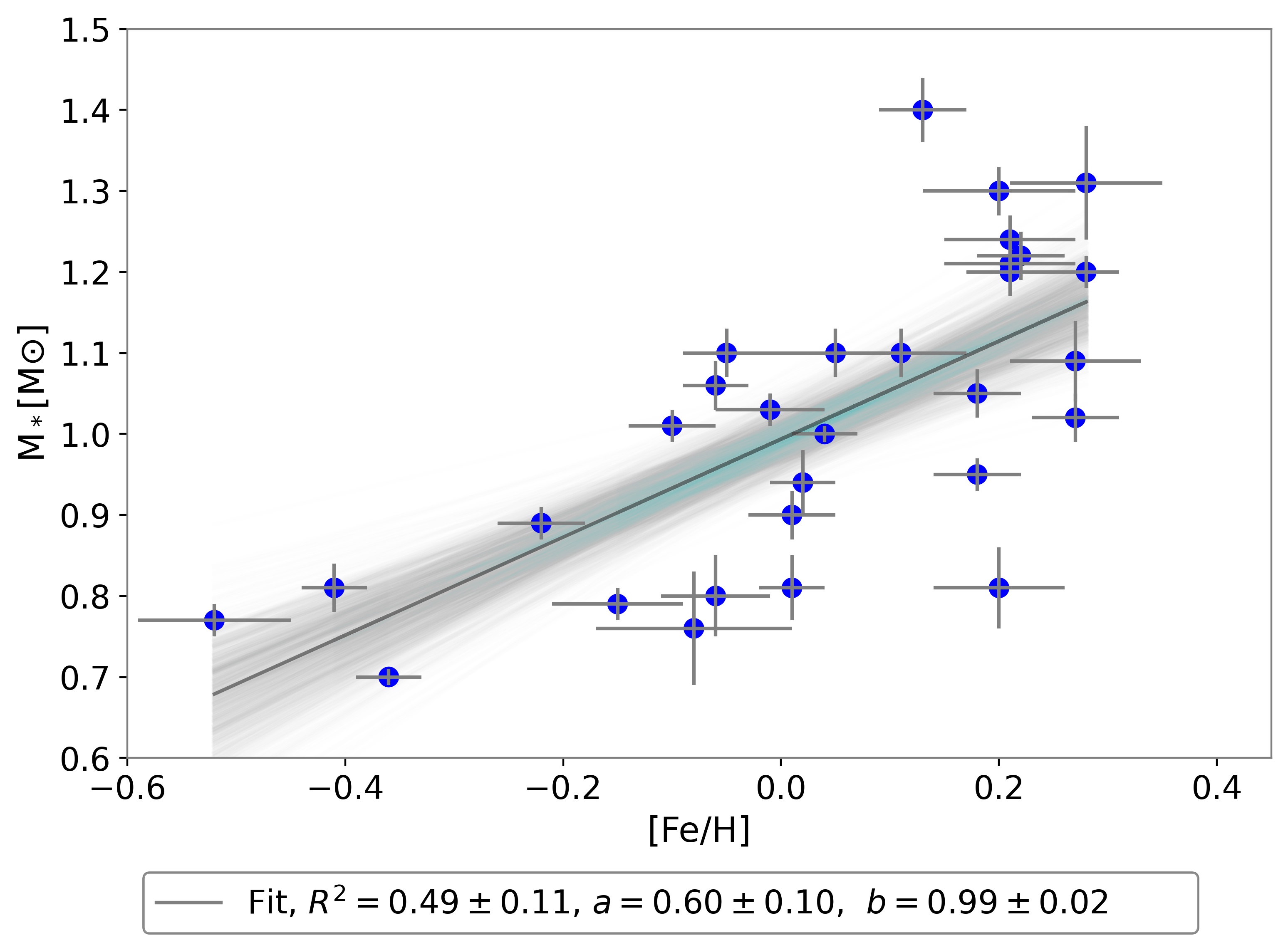}\par 
   \caption{Stellar mass of the DDP stars of Table~\ref{t:table1} versus [Fe/H]. Blue dots show the observational data and the solid black line shows the best linear regression (y = ax + b), calculated using bootstrapping method weighted by the errors of metallicity, the
parameter values are shown in the box below the figure. The shaded blue area shows the 68 per cent confidence band of the bootstrapping fit. }
   \label{fig3}
    \end{figure}

In Fig.~\ref{fig3}, we present the stellar masses of stars hosting planets and debris disks as a function of the metallicity [Fe/H] for the whole set of DDP objects presented in Table \ref{t:table1}. The stellar mass is a good proxy of the protoplanetary disk properties, as demonstrated by, for instance, \citet{Pascucci16}. As can be seen, a relatively good positive gradient is obtained for the less massive DDP stars with the lowest metallicities. This result is in good agreement with those found in previous studies \citep{Gonzalez1997,FischerValenti2005}, where it was concluded that high metallicity favours the formation of planets around more massive stars, while planets form preferentially in lower mass stars when the metallicity is low. This is a subject that was also recently reviewed by \citet{Adibekyan2019}. After a deep analysis of other possible correlations, we did not find the expected clear correlation between the total accumulated mass in planets and the [Fe/H]. The lack of a correlation may be a result of the complexity of the planet formation process that can vary from star to star or to the incompleteness of the sample.\\

\begin{figure}
 \centering
 \includegraphics[width=9cm]{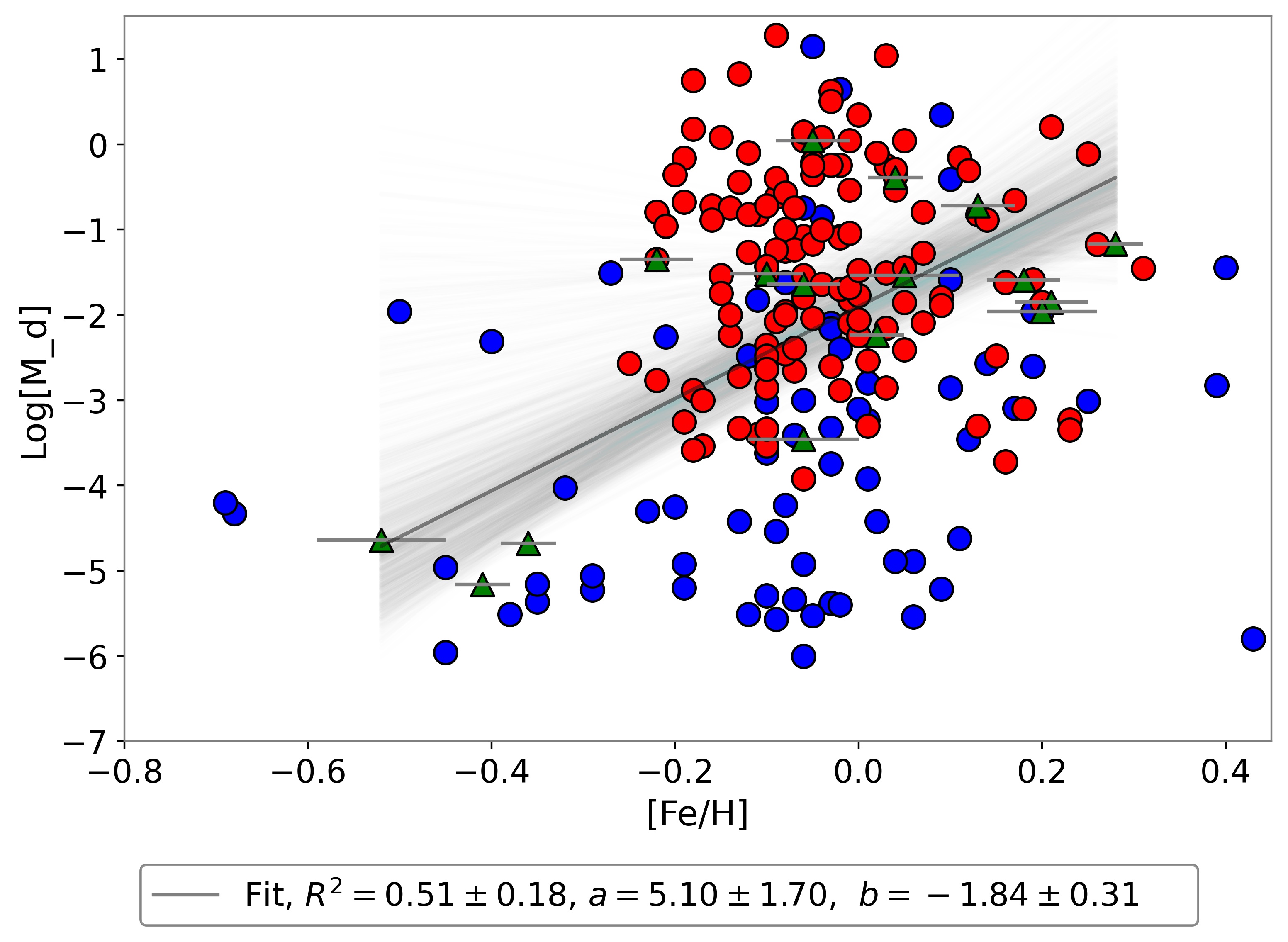}\par 
 \caption{In this figure, we show the stellar disk masses of DD stars (in
lunar masses) of part of the \citet{Chen2014} catalog, as a  function of [Fe/H].
Red points are the values of the masses given in the second disk distance model
($M_{\rm d2}$) of this catalog. Blue points are the values corresponding to the single
distance model ($M_{\rm d}$). Green triangles are the DDP stars of Table~\ref{t:table1} with $M_{\rm d}$ values
both from Chen’s catalog and the three computed in the current work (see Sect.~\ref{sec:diskmass}). The green solid line 
shows the best linear regression (y = ax + b), calculated using a bootstrapping method weighted by the errors on metallicity for all the DDP stars with known $M_{\rm d}$ -- but excluding the HAHM objects that have a different origin than those studied here (see discussion in Sect.~\ref{sec:conc}). The 
parameter values are shown in the box below the figure. The shaded blue area shows the 68 per cent confidence band of the bootstrapping fit.
} 
\label{fig4}
\end{figure}

Focusing now on the debris disks properties, we also tried to find correlations between the debris disk mass and the metallicity of the host star, as we did in CH19 but this time we have included information on their origin in the thin or  thick Galactic-disk populations. In this study, we used all the currently known DDP stars (see Table~\ref{t:table1}), but we only analyzed the ones with the debris disk mass obtained from the infrared emission at 70\,$\mu$m with Herschel \citep{Chen2014}, as well as the three for which it was obtained  (see Sect.~\ref{sec:diskmass}). Emission at this wavelength is produced by dust grains with diameters smaller than a millimetre and can radiate up to distances of some tens of AU from the central star. However, we note (and as is discussed in CH19) that a similar result on the correlation between dust mass and [Fe/H] has been detected for larger grain sizes up to more than mm at wavelength of 850 $\mu$m, which has  allowed researchers to reach much larger distances from the central star. We consider that at the light of these conclusions our results obtained from the 70\,$\mu$m observations can be extrapolated to match the ones that cover disk sizes from one to ten times the size of the Kuiper Belt. The [Fe/H] values have been obtained by cross-matching the results from  \citet{Chen2014} with those of \citet{Gaspar2016}. We show our results for the DDP stars in Fig.~\ref{fig4} as green triangles. We also include a regression line that shows  how we reproduce the gradient observed by CH19 for small grains. It is crucial to emphasize here that the gradient is only recovered if we include the three thick-disk stars presented in the previous sections, that is, we show how the gradient is caused by the evolution of metallicity in the ISM from thick disk to thin disk formation.

\subsection{Metallicity gradient of DD stars}\label{subsec:metalsDD}

In addition to the DDP stars in Fig.~\ref{fig4}, we also show a large fraction of the DD stars presented in the \citet{Chen2014} catalog (red and blue symbols). Here, we further push  the analysis of correlations between metallicity and parameters of the debris disks around stars presented in CH19 and references therein. In these previous works, the authors concluded that there is no correlation between the dust properties and metallicity. In these studies, the authors used IR-based parameters, such as as $L_{IR}/{L_*}$, as well  as  debris disk masses $M_{\rm d}$  (e.g., CH19). We state that the lack of correlation was due to the fact that the authors only analyzed stars from the thin disk, but if we include stars from other galactic components (i.e., thick-disk stars),  correlations do indeed arise. This is illustrated by Fig.~\ref{fig4}, where we show that the distribution of stars with values of [Fe/H] $< -0.2$ (mostly thick-disk stars) is different from what we see for higher metallicities. In particular, the few DD stars with low metallicities contain mostly low- and very low-mass debris disks, and they may also belong to the thick disk (see Table~\ref{t:table1}). Otherwise, high-metallicity, thin-disk stars have enough material in their surroundings to produce a very rich variety of debris disks with a variety of morphologies and sizes -- effectively  eliminating any correlations with [Fe/H].

\subsection{Hunting for the oldest DD, DDP, and
CP stars in the Galaxy}\label{subsec:primitiveDD}

Recently, thanks to the publication of massive databases of stellar atmospheric parameters and kinematics, a new picture has emerged concerning the formation timescale  of our Galaxy  \citep{xiang2022}. In particular, the history of its star formation has been rewritten in the last few years. An important result regarding the formation of what is called the thick disk was obtained by analyzing 250000 subgiant stars that are considered to be one of the best age indicators due to their short life. These authors presented results to suggest that the old thick disk began its formation   $\sim$13 Gyr ago, that is, only 0.8\,Gyr after the Big Bang, with the star formation peak occurring at 11\,Gyr, which coincides with the merger of our galaxy with the satellite galaxy Gaia-Sausage-Enceladus. This result is compatible with others based only on
abundances \citep{Haywood2016,MaozGraur2017} or on analyses of color-magnitude diagrams \citep{Mor2019}. This peak was followed by a quenching of the star formation, a result that coincides with post-merger predictions from theoretical models. The quenched star formation lasted $\sim$2\,Gyr, from 3 to 5\,Gyr. A few of the DD, DDP, and CP stars contained in Table~\ref{t:table1} were born in this period, so we can consider them as participants in the formation process of the thick disk. Authors such as  \citet{Sheehan2010}, \citet{Shchekinov2013}, \citet{Campante2015} and \citet{Hasegawa2014} have explored the very-low-metallicity regime for stars in past and current large databases, looking for the first planetary systems formed in the Galaxy. In these
works, the authors only looked for CP stars, whereas here we are looking  for the most primitive DDP stars and some DD stars for the
first time. All of our candidates are listed in Table~\ref{t:table1} and the DD and DDP are also represented in Fig.~\ref{fig4}.\\ 

Now we adopt the following conditions in order to select the best candidates for the oldest DD, DDP, and CP main-sequence stars: a) to have an age determination with a small error and with a value higher than 8\,Gyr; b) to be a member of the thick disk, as discussed in Sect.~\ref{sec:data}; c) to have a very small lithium abundance indicating
an expected and very high depletion of this element. The DD star HD\,110897 with A(Li) = 1.94 and the CP star HD\,114729 with A(Li) = 1.88 (which are even considered as thick-disk stars in Table~\ref{t:table1}) were not labeled as primitive due to their high A(Li) values. All these results are presented in Table~\ref{t:table3}, where two DD stars, five CP and three DDP are shown as likely to have been formed during the maximum at 11\,Gyr as well as in the following quenching of star formation of the thick disk.\\

Among the CP stars we considered as the “most primitive,” the results from the HD\,189567 and the HIP\,94931 (also known as K-444) stars are especially relevant as the age estimations are precise and they have very low metallicity and A(Li), suggesting that may be the oldest stars with known planets. A similar result but for the DDP stars is the one obtained from  the HD\,20794 star. This star has an age that exceeds by several Gyrs the one of the other two candidates. The mass of the debris disk around this star ($M_{\rm d}$) is much lower than the one in all other candidates (see Sect.~\ref{sec:data}). In light of these results, we conclude that HD\,20794 is the oldest DDP star known in the Galaxy. 
Finally, we also notice  that, as expected, the planetary masses of all the thick-disk primitive stars mentioned here (both CP and DDP) are very small, with the sole exception of HD\,37124, which has a total planetary mass of 2.02 $M_{\rm Jup}$. This star was classified as transitory disk star, as it should belong to the thin disk according to the stellar kinematics criteria -- however, it would be attributed to the thick disk based on the analysis of  its [Ti/Fe] versus [Fe/H] measurement (see Table~\ref{t:table1} and the discussion in Sect.~\ref{sec:data}). We have not reached any conclusion in this work to answer why it hosts these large planets and we state that new observations are needed to better understand the origin and properties of this object.

\begin{center}
\begin{table}
\caption{Properties of the old and low-[Fe/H] DD, DDP,  and CP stars in our sample. 
 Col. 1: HD star name. Col. 2: group defined in Section 1:  CP, DD, and DDP. Col. 3: spectral type. Col. 4: age. Col. 5: metallicity and
 Col. 6: lithium abundance, extracted from Table~\ref{t:table1}.}
\label{t:table3}
\scalebox{0.85}{
\begin{tabular}{lccccc}
\hline
\hline
 Star    & Object     & ST &   Age   & [Fe/H]  & A(Li)   \\
         &            &    & (Gyr)   & dex     &              \\
   \hline
 HD 101259  & DD     & G6/G8~V  & 6.8$^{+4.8}_{-3.9}$      & -0.746 &  0.8     \\
 HD 213941  & DD     & G8       & 12.811 $\pm$ 0.536                  & -0.46 &  0.34         \\
 HD 6434    & CP     & G2/3~V   & 9.55 $\pm$ 2.95                     & -0.61 &  0.84         \\
 HD 37124   & CP     & G4~IV/V  & 11.10 $\pm$ 1.7                     & -0.43 &  0.49          \\
 HD 136352  & CP     & G3/G5~V  & 9.6 $\pm$ 1.8                       & -0.28 &  <0.86        \\
 HD 189567  & CP     & G2~V     & 11.0 $\pm$ 0.5                      & -0.24 &  0.86         \\
 HIP 94931  & CP     & K0~V     & 11.54 $\pm$ 0.99                    & -0.37 &  -0.34         \\
 HD 10700   & DDP    & G8~V     & 8.2 $\pm$ 3.2                       & -0.52 &  <0.42          \\
 HD 20794   & DDP    & G8~V     & 11.4 $\pm$ 0.3                      & -0.41 &  <0.58          \\
 HD 40307   & DDP    & K3~V     & 6.0 $\pm$ 4.1                       & -0.36 &  < 0.22         \\
\hline 
\end{tabular}
}
\end{table}
\end{center}
 
\begin{figure*}[!]
\centering
  \subfigure[\hbox{ A(Li) versus [Fe/H]}]{\includegraphics{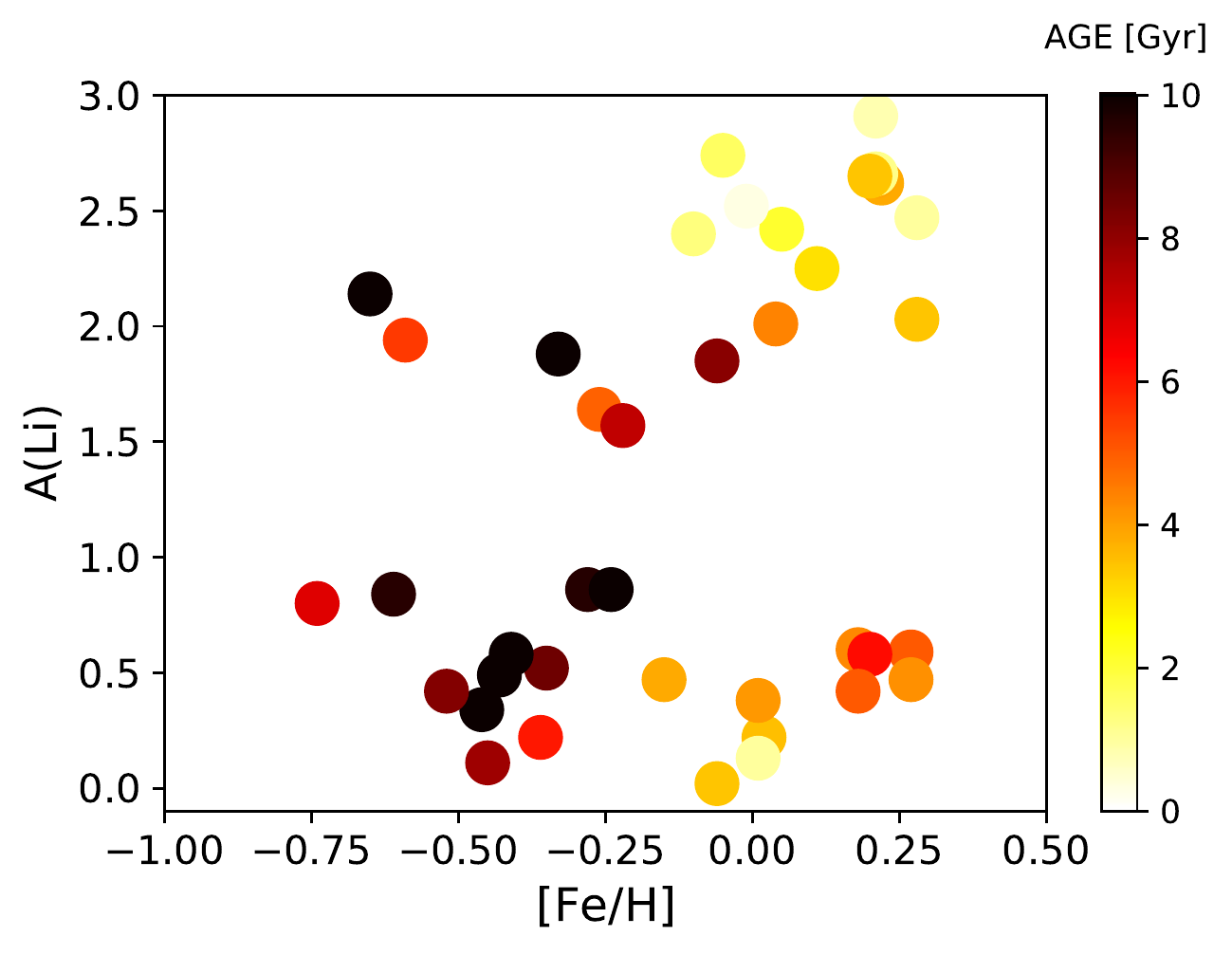}}
  \hfill
    \subfigure[\hbox{A(Li) versus [Fe/H]}]{\includegraphics{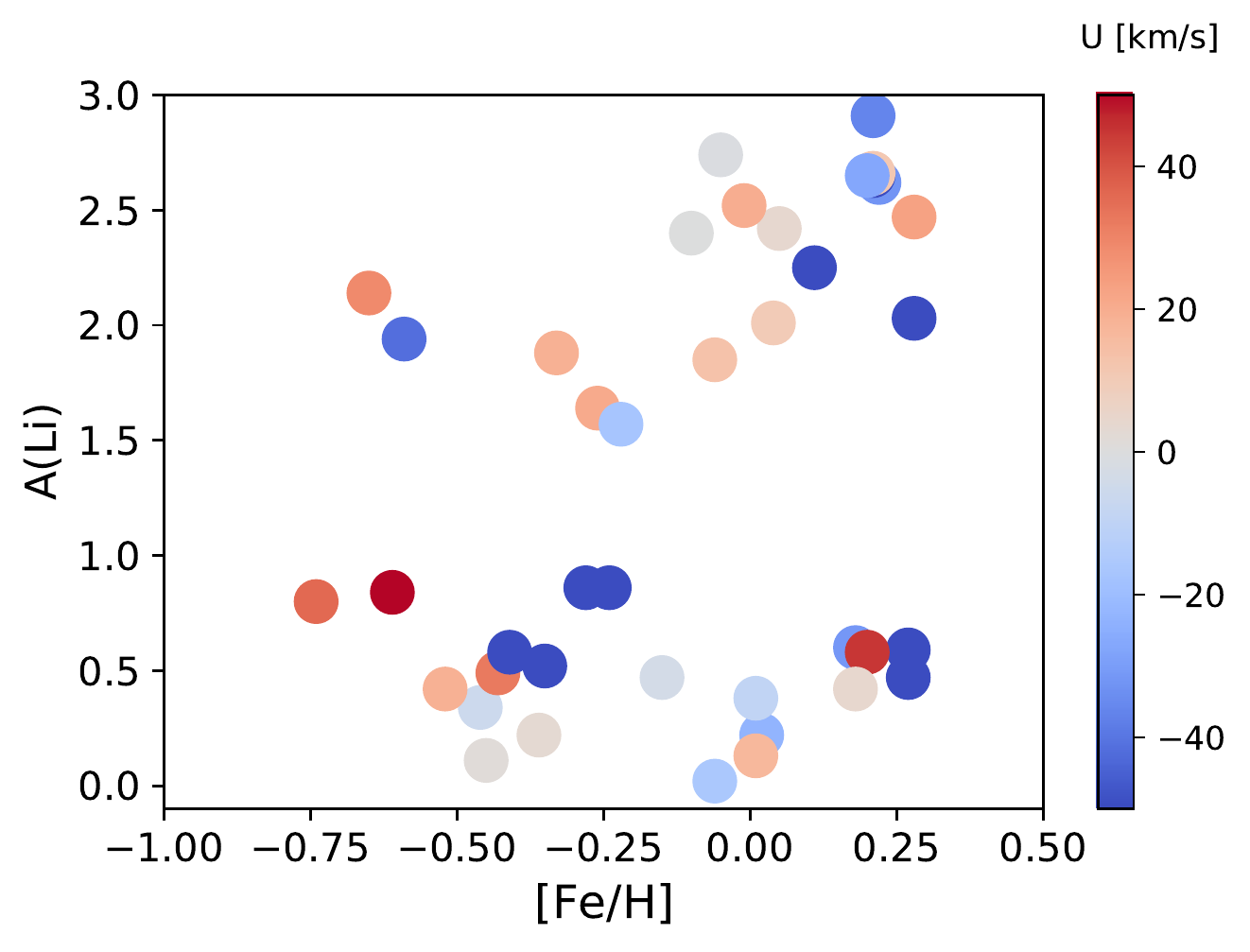}}
  \hfill
\caption{ A(Li)  as a function of [Fe/H] for all stars listed in Table 1,  shown as two different
branches. The top branch, with A(Li)  $>$ 1.0,
represents the expected Li evolution in the Galactic thin disk. The
bottom branch, with very depleted Li values less than 1.0, contains a
mixture of thick-disk stars in the low-metallicity domain with [Fe/H] $<$ -0.2 and thin-disk stars with larger metallicities values
up to 0.28. Among the thin-disk, Li-depleted stars, with  -0.15 $<$[Fe/H]$<$ 0.01,  we include the HAHM stars discussed in Sect. 2.1. The corresponding ages (a) and the U Galactic radial velocities (b)  are shown in vertical color  scales.
 }
 \label{fig5}
\end{figure*}  
 
\subsection{ Lithium properties of the DD, DDP and CP stars}\label{subsec:lithium}

The lithium abundance and its evolution inside FGK dwarf stars have been widely studied in the past. In particular, many authors have shown that despite its fragility, the A(Li) abundance can be used as a tracer of metallicity in young stars \citep{Rebolo88a,Rebolo88b}. Stars inherit the metallicity and lithium abundance of the interstellar medium where they were born and via numerous mechanisms inside and outside, stars continue to grow as the Universe gets older \citep{RocaFabrega21}. If no other agent is present, the A(Li) should increase at a similar rate as the metallicity. However, observations have shown that some metal-rich dwarf stars in the solar neighbourhood ([Fe/H] $>$ 0.15) have low A(Li). This can be easily explained if a lithium destruction mechanism is present inside stars \citep{Eggenberger} and if these metal-rich dwarf stars are old enough to allow this mechanism to work until it depletes most of the lithium \citep{RocaFabrega21}. The main problem with this hypothesis is that the metal-rich stars in the solar neighbourhood are usually the youngest. In \cite{Dantas2022}, the authors present a detailed discussion of the  a suggestion from \citet{Guiglion2019} aimed at solving this problem. \citet{Guiglion2019} proposed that these metal-rich dwarf stars are indeed old stars but they happen to come from the inner regions of the Galaxy, which are more metal-rich than the solar neighbourhood. These stars thus depleted their Li in their journey to their current positions.

In this context, as our sample of stars covers a broad metallicity range, we study whether some of the objects presented in Table~\ref{t:table1} show peculiar A(Li) values. In Figure~\ref{fig5}, we show the A(Li) versus [Fe/H] colored by age (left) and galactocentric radial velocity (right) and we note the presence of two branches: one lithium-rich (A(Li)>1.5) and one lithium-poor (A(Li)<1.0). Interestingly, although most lithium-poor stars are also metal deficient, five stars have low metallicity ([Fe/H]<-0.2) and high A(Li). Also, we see that five stars with super-solar metallicity have very low A(Li). If we focus on the five metal-rich stars with low A(Li), we can try to test the Guiglion hypothesis by using the stellar kinematics provided by the Gaia satellite and ages presented in Table~\ref{t:table1} (see color bars in Figure~\ref{fig5}). After computing the mean ages of the metal-rich stars in the two branches, namely, A(Li)>1.5 and A(Li)<1.0, we find that the former is much younger than the latter ($\sim$ 2 Gyr  versus $\sim$ 5 Gyr); this is also evident when comparing the colors in the left panel of Figure 5. This result is in agreement with the Guiglion scenario. Now, if we examine the Galactic radial velocity of these stars ($U$), which may be a tracer of radial migration, we see that three out of five of the low-A(Li) stars show high outward (negative) radial velocities, two of them with values higher than -50\,kms$^{-1}$ and one with high inward velocity (positive) that is the eldest (HD\,108874). Stars in the high-A(Li) branch show radial velocities lower than 50\,kms$^{-1}$ (absolute), with the only exception being that of the eldest star (HD\,52265), which exhibits a high outward velocity.

These results show that given our poor statistics, we cannot make a strong statement about the validity of the Guiglion hypothesis. However, we can point out that at least three low-A(Li) objects  may be candidates for the status of migratory stars from the Galactic central regions. Also, we can confirm that stars in the low-A(Li) branch are older than the ones in the high-A(Li). Finally, it is worth mentioning that using a larger sample of stars and a different approach, \cite{Dantas2022} gave an initial (but not final) confirmation of the Guiglion hypothesis.

\section{ Discussion and conclusions}\label{sec:conc}

From the thousands of stars with confirmed planets (CP), we detected second-generation debris disks, that is, Kuiper belt-like structures in only a few (i.e., less
than 40). The lack of such structures remains an open question. 
Many hypotheses can be proposed, starting  from a possible observational bias or a DDP short lifetime (i.e., fast transition from DDP to CP) to a mechanism that stabilises the planetary system and does not allow for the formation
of this second-generation of cold debris disks in most stars \citep{Najita2022}. However, none of these hypotheses have been confirmed or rejected thus far.\\

Some recent results on the DDP nature show that the presence of debris disks around stars with confirmed planets is independent of the masses of the planets and of its age, as well as of its metallicity \citep[e.g.,][]{Montesinos2016} -- this is a result that also applies to DD stars without confirmed planets. We show in Sect.~\ref{sec:data} that, in fact, the large majority of these objects are members of the young and metal-rich Galactic thin disk and we hypothesize that this may be the cause of the lack of correlation with metallicity found in previous works, such as that of \citet{Maldonado2012}. In fact, many authors have shown that in a metal-rich thin disk, the broad distribution of dust grain sizes, each with a different metallicity dependence, can undermine any possible correlation (see e.g., CH19 and references therein). What is even more interesting is the fact that we have detected three DDP stars belonging to the old and metal-poor Galactic thick disk. One of them, HD\,20794, was presumably formed at the maximum stellar formation rate of the thick disk 11\,Gyr ago (see Sect.~\ref{subsec:primitiveDD}). This epoch, as proposed by \cite{xiang2022}, coincides with that of the merger of the Galaxy with the Gaia-Sausage-Enceladus. The other two DDP stars (HD\,10700 and HD\,40307) would belong to the following quenching of star formation that lasted down to 7-9\,Gyr ago. These three stars were not included in the former work by \citet{Chen2014} and the mass of their debris disk was not computed. In this work, we carry out that computation and we show that the detection of DDP stars in the thick disk naturally generates a correlation between the debris disk mass and the metallicity, with the metal-poor, thick-disk stars being the ones that are shown to host  debris disks of lower masses (see Sect.~\ref{subsec:metalDDP}). We note here that although these results were obtained using data for the debris disks from the 70\,$\mu$m emission -- that is, from dust grains with diameters smaller than a mm and extending to distances of several tens of AU from the central star -- in CH19, we showed that this debris disk mass-metallicity correlation is maintained when larger than a mm grains are considered. With emissions at 850\,$\mu$m the gradient can be conserved up to much larger distances on the order of one to ten times the Solar System Kuiper belt radius.\\

Regarding the debris disk mass of the three thick-disk DDPs, we found that when they are measured at 70\,$\mu$m, all of them are very small. This result agrees with a plausible correlation between metallicity and the DD mass (see Fig.~\ref{fig4}), meaning that very old DD stars that formed during a low-metal regime in the Galactic thick disk would be expected to host low-mass debris disks. These masses are several orders of magnitude lower than the mass of the Kuiper belt (KB) of our Solar System. For the DDP star HD\,20794, which is the most primitive one, its debris disk mass is even smaller. It is only the disk of the star Tau Cet that presents an anomaly. \citet{Greaves2004a} obtained a mass for this debris disk of $\sim$\,1.2 Earth masses, measured at 850\,$\mu$m. The origin of such a massive disk is unclear and whether it is a transient phenomena or not and how frequent can be these kind of events is still under debate. We assert that it is only a scenario of a transient replenishment of the debris disks that can fit this system within the general picture of  debris disks that are slowly decaying in line with their stellar ages.

In our analysis of the CP, DDP, and DD stars, we have found four stars that fall within the regime of the so known as high-alpha high-metallicity (HAHM) stars. The existence of this kind of star has been discussed by \citet{Adibekyan2012}, but the authors have not succeeded in establishing a final conclusion with regard to their origin. The problem revolves around the fact that the high $\alpha$ abundances of these stars would indicate they belong to an old population (e.g., the thick disk), but their kinematics suggest they should belong to the thin disk. In this work, the four HAHM stars are young (even with larger uncertainties) and clearly belong to the thin disk when applying the kinematic criteria, but they do show a high [$\alpha$/Fe] (where we use Ti as a proxy for the $\alpha$-elements). We speculate that they could have been enriched with alpha elements during their evolution by direct accretion -- or, more probably, that they were born in non well-mixed clouds of low-[Fe/H] gas \citep{vandenHoek1997} or in low-[Fe/H] gas accreted in a recent extragalactic or circumgalactic gas infall \citep{Chiappini2015}. However, this question requires  further investigation to reach a firmer conclusion.\\
 
In this work, we also consider the main lithium properties of our stars, mainly investigating  a few super-metal-rich stars that are highly lithium-depleted.
We made an initial approach, using our ages and Galactic kinematical data, and we found indications supporting the scenario proposed by other authors, namely, these stars are migrating from internal metal-rich regions of the Galaxy and depleting their lithium along the journey to their current positions.

Finally,  as a complementary result to this research, we found some DD, DDP, and CP stars that could be considered to be the oldest objects of this kind in the thick disk. The final results of this search for primitive stars can be found in Table~\ref{t:table3}. where we show two DD stars, four CP stars, and three DDP stars with old ages and very small total planetary masses. This is a result that is in agreement with the assumption that these stars were born into a very-low-metal regime. In particular, we note that star HD 20794, thanks to the high precision on its age estimation and to  extremely small disk mass, can be considered the most primitive known system containing debris disk and planets in the Galaxy.

\begin{acknowledgements}
SFR acknowledges support from a Comunidad de Madrid postdoctoral fellowship under grant number 2017-T2/TIC-5592. His work has been supported by the Madrid Government under the Multiannual Agreement with UCM in the line Program to Stimulate Research for Young Doctors in the context of the VPRICIT under grant number PR65/19-22462. SRF also acknowledges financial support from MINECO under grant number AYA2017-90589-REDT, RTI2018-096188-B-I00, and S2018/NMT-429. FLA and PC acknowledge the support from the Faculty of the European Space Astronomy Centre (ESAC), under funding reference number ESA-SCI-SC-LE-059.
PC acknowledges financial support from the Government of Comunidad Autónoma de Madrid (Spain) via postdoctoral grant `Atracción de Talento Investigador' 2019-T2/TIC-14760.
CC acknowledges financial support from the Agencia Estatal de Investigación of the Ministerio de Ciencia y Universidades through project AYA2016-79425-C3-1/2/3-P.
We sincerely appreciate  the constructive comments of an anonymous referee.

\end{acknowledgements}

\begin{appendix}

\section{Calculations of  $M_{\rm d}$}\label{app:1}

For the calculations to obtain  $M_{\rm d}$, we used the relation (11) in \cite{Chen2014}: 

\begin{equation}
 M_{\rm d} = \frac{3\pi}{16}\frac{L_{IR}}{L_* }\rho R^2<a>
,\end{equation}

 where  $\rho$ is the mass density, equal to 3.3 $g cm^{-3}$, $R$ is the radius or the dust distance from the
central star, and <a> is the radius of emitting grains equal to 5/3 $a$(min)
in $\mu$m. This last value depends on the spectral type of
the star, obtained from the tables
presented in \cite{Chen2014}. Also, $L_{IR}/L_*$ is the ratio  
between the luminosity of the dust emitting zone of
the disk and the stellar luminosity. We present in Table~\ref{t:appendix}
the parameters included in the calculation of
$M_{\rm d}$ and  their references.

\begin{center}
\begin{table}[h]
\caption{Parameters of  $M_{\rm d}$ calculation}
\label{t:appendix}
\scalebox{0.90}{
\begin{tabular}{lcccccc}
\hline
\hline
Star  & SpType     & $a$(min)   &   $L_{IR}/{L_\star}$ &   R  & Ref.& $M_{\rm d}$      \\
      &            &   $\mu$m   &                      & (AU) &     & ($M_{\rm (Moon)}$)         \\\hline 
  
HD20794     & G8V    & 0.7    &  1.6x$10^{-3}$      & 15     & a & 7.09$\,\times\,10^{-6}$   \\
HD10700     & G8V    &  0.7   &    6.1x$10^{-6}$    &   14  & a & 2.30$\,\times\,10^{-5}$ \\  
HD40307     & K3V    & 0.3    &    4.3x$10^{-6}$     &  24   & b  & 2.10$\,\times\,10^{-5}$             \\  
\\
\hline 
\end{tabular}
}
\tablefoot{ a) \citet{Sibthorpe2018}, b) \citet{Marshall2014}}
\end{table}
\end{center}
 
\end{appendix}

\end{document}